\documentclass[prl,aps,twocolumn,showpacs,preprintnumbers,amsmath,amssymb, superscriptaddress]{revtex4-2}

\usepackage[makeroom]{cancel}

\usepackage{amsmath}    
\usepackage{amssymb}
\usepackage{graphicx}   
\usepackage{verbatim}   
\usepackage{color}      
\usepackage{hyperref}   
\usepackage[normalem]{ulem}
\usepackage{natbib}
\usepackage{fixmath}
\usepackage{enumitem}
\usepackage{empheq}

\hypersetup{colorlinks,linkcolor=blue,urlcolor=blue,citecolor=blue}

\definecolor{darkgreen}{rgb}{0,0.5,0}
\definecolor{purple}{rgb}{0.35,0,0.35}
\definecolor{orange}{rgb}{1,0.5,0}
\definecolor{darkred}{rgb}{.7,0,0}
\definecolor{darkblue}{rgb}{0,0,.3}
\definecolor{grey}{rgb}{.6,.6,.6}
\definecolor{dimgreen}{rgb}{0.2,0.6,0.1}

\definecolor{darkgreen}{rgb}{0,0.5,0}

\begin{document}

\title{Full counting statistics in the many-body Hatano-Nelson model}

\author{Bal\'azs D\'ora}
\email{dora.balazs@ttk.bme.hu}
\affiliation{MTA-BME Lend\"ulet Topology and Correlation Research Group,
Budapest University of Technology and Economics, 1521 Budapest, Hungary}
\affiliation{Department of Theoretical Physics, Budapest University of Technology and Economics, 1521 Budapest, Hungary}
\author{C\u{a}t\u{a}lin Pa\c{s}cu Moca}
\affiliation{MTA-BME Quantum Dynamics and Correlations Research Group, Institute of Physics, Budapest University of Technology and Economics, 1521 Budapest, Hungary}
\affiliation{Department  of  Physics,  University  of  Oradea,  410087,  Oradea,  Romania}

\date{\today}

\begin{abstract}
We study non-hermitian many-body physics in the interacting Hatano-Nelson model with open boundary condition. 
The violation of reciprocity, resulting from  an imaginary vector potential, induces the 
non-hermitian skin-effect and causes exponential localization for all single particle eigenfunctions in the non-interacting limit.
Nevertheless, the density profile of the interacting system becomes only slightly tilted relative to the average filling.
The Friedel oscillations exhibit a beating pattern due to the modification of the Fermi wavenumber.
%
%
The probability distribution of particles over any finite interval is the  normal distribution, whose mean scales
with the imaginary vector potential and the variance is symmetric to the center of the chain. 
This is confirmed by several numerically exact methods even for relatively small systems.
These features are expected to be generic not only for fermions, which naturally repel each other due to Pauli's exclusion principle,
but for interacting bosons as well.
Our findings indicate that many-body effects can significantly alter and conceal the single particle properties and the skin effect
in non-hermitian systems.
\end{abstract}

\maketitle

\paragraph{Introduction.}

In the theory of probability,  a classical random variable is often successfully characterized by a few of its first moments,
 and the powerful laws that follow, such as the central limit theorem\cite{fischer},
 have found numerous applications in diverse areas of science\cite{touchette}.
In many cases, however, this approach ceases to be satisfactory, and
the full probability distribution function  is needed as it reveals salient features about the random variable\cite{bertin}.
The quantum world is not much different in this respect from the classical one. Already simple expectation values of physical quantities
often display rather complex behavior, and their complete understanding requires lifelong effort, such as e.g. the phase diagram of
the  Hubbard model\cite{arovas}. 
Although they are often difficult to access, higher moments of the observables 
encapsulate unique information about non-local, multi-point correlators and 
entanglement and contain much more information.
Finding all these moments is essentially equivalent to determining the entire distribution 
function of the quantity of interest, i.e. the characteristic function of full counting statistics\cite{rmptalkner,ashida2018,levitov,hofferberthnatphys,gritsev,gring,herce}.

Recent years have witnessed an explosion of interest towards non-hermitian quantum systems\cite{Bergholtz2021,ashidareview,rotter,ElGanainy2018}. 
The ensuing physics often arises
from considering open quantum system interacting with their environment \'a la Lindblad and continuous monitoring together with postselection\cite{daley,carmichael}, giving rise
to many unexpected phenomena with no obvious counterpart in a hermitian setting. One unique feature of non-hermitian systems 
is the anomalous localization of all eigenstates coined as the non-Hermitian skin effect\cite{lee2016,yao2018,kunst2018} when all 
single particle eigenstates become exponentially localized at the boundaries of the system. 
A paradigmatic model associated to this is the Hatano-Nelson\cite{hatanonelson1,hatanonelson2} model, where 
the breakdown of reciprocity, i.e. asymmetric hoppings induces the skin effect.
While this occurs at the single particle level, several studies addressed the fate of the non-hermitian  skin-effect 
in a many-body setting, including numerics\cite{zhang2022,alsallom,lee2020,hamazaki} as well as Bethe ansatz\cite{mao2022}. 

Here we go one step further and combine the above two concepts, namely full counting statistics and non-hermitian physics, 
by analyzing the probability distribution of particle number over a finite interval for the interacting  Hatano-Nelson chain.
By using bosonization, we solve the low energy effective theory exactly and obtain analytical results for the real space density profile as well as
the characteristic function of the particle density over a finite interval. 
We find that many-body physics suppresses significantly the non-hermitian skin effect
both for fermions and bosons. The Friedel oscillations exhibit beating pattern as the Fermi wavenumber gets modified by the
presence of the imaginary vector potential. 
In spite of the skin-effect at the single particle level, 
the probability distribution of particle density over a finite interval remains normal and depends on the location of the interval within the open chain.
The mean value of the normal distribution scales with the imaginary vector potential while the variance becomes independent from it. 
Our analytical findings are corroborated  by several numerically exact methods even for relatively small systems.

\paragraph{Hamiltonian.}
The Hatano-Nelson model\cite{hatanonelson1,hatanonelson2} consists of fermions hopping in one dimension in the presence of an imaginary vector potential. The interacting 
many-body version of the Hamiltonian is
\begin{gather}
H=\sum_{n=1}^{N-1} \frac J2 \exp(ah)c^+_nc_{n+1}+\frac J2 \exp(-ah) c^+_{n+1}c_n+\nonumber \\
+U c^+_nc_{n}c^+_{n+1}c_{n+1},
\label{hamiltontb}
\end{gather}
where $J$ is the uniform hopping, $h$ is the constant imaginary vector potential and $a$ represents the lattice constant, 
$N$ is the total number of lattice sites and we consider
open boundary condition (OBC), $U$ represents the the nearest-neighbour interaction between particles.
We consider half filling with $N/2$ fermions populating the lattice. 
 The model is PT-symmetric\cite{bender2007} and possesses a real spectrum for OBC, and the minimal energy 
configuration is the ground state with many-body wavefunction $|\Psi\rangle$.
Due to OBC, no current is expected to flow in the system but the real space density profile is expected to be inhomogeneous due to the imaginary vector potential.
For periodic boundary condition, a finite persistent current circulates in the system\cite{Affleck2004,Hofstetter2004} but the real space density profile is homogeneous.

\begin{figure}[h!]
\centering
\includegraphics[width=7cm]{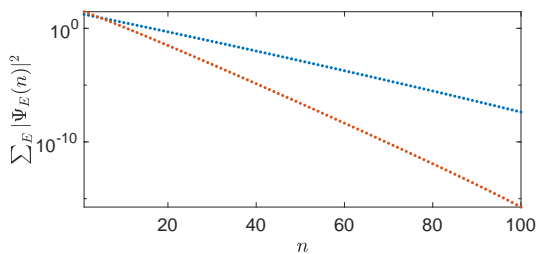}
\caption{Sum of absolute squares of amplitudes per site\cite{Bergholtz2021} of all right eigenstates of the non-interacting $H$
 with $U=0$, $N=100$ and OBC with $ha=0.1$ (upper blue) and 0.2 (lower red), exhibiting the non-hermitian skin effect.
}  
\label{hnskin}
\end{figure}

In the non-interacting, $U=0$ limit, the model can be diagonalized and the single particle eigenfunctions, $\Psi_E(n)$ for eigenenergy 
$E$ can be obtained. With OBC, these are localized to one end 
of the system (dictated by the sign of $h$) as a manifestation of the non-hermitian skin effect. This is shown in Fig. \ref{hnskin}.

Upon going to the continuum limit and bosonizing the theory~\cite{giamarchi,cazalillaboson,nersesyan}, the effective low energy Hamiltonian reads as
\begin{gather}
H=\int_0^L \frac{dx}{2\pi} v\left[K(\pi\Pi(x)-i\pi h)^2+\frac 1K (\partial_x\phi(x))^2\right],
\label{hamboson}
\end{gather}
where $\Pi$ and $\phi$ are dual fields satisfying $[\phi(x_1),\Pi(x_2)]=i\delta(x_1-x_2)$.
This Hamiltonian is brought into conventional hermitian Gaussian form by applying a similarity transformation, which eliminates the $ih$ term using $S^{-1}HS$ with 
$S=\exp(-\frac h\pi \int_0^L\phi(x')dx')$.
The resulting Hamiltonian can be brought to diagonal form after introducing canonical bosonic fields\cite{giamarchi} as 
\begin{gather}
H=\sum_{q>0}\omega(q)b^\dagger_qb_q,
\label{hboson}
\end{gather}
 and the long wavelength part of the local charge density is $\partial_x \phi(x)/\pi$
with
\begin{gather}
 \phi(x)=i\sum_{q>0}\sqrt{\frac{\pi K}{qL}}\sin(qx)\left[b_q-b^\dagger_q\right]
\label{thetax}
\end{gather}
for OBC
and $K$ the LL parameter\cite{giamarchi} (which carries all the non-perturbative effects of interaction $U$ from Eq. \eqref{hamiltontb}) 
and $\omega(q)=vq$ with $v$ the Fermi velocity in the interacting systems and $q=l\pi/L$ with $l=1,2,3\dots$.
The ground state of Eq. \eqref{hboson} is the vacuum state $|0\rangle$, and the ground state of the original non-hermitian model is obtained as
$|\Psi\rangle=S|0\rangle/\sqrt{\langle 0|S^2| 0\rangle}$. The very fact that we managed to manipulate the low energy effective theory into Eq. \eqref{hboson} indicates that the interacting
Hatano-Nelson model indeed forms a Luttinger liquid with collective bosonic excitations in much the same way as for hermitian systems\cite{giamarchi}.

Any expectation value of an operator $\mathcal O$ is evaluated as
\begin{gather}
\langle \Psi|{\mathcal O}|\Psi\rangle=\frac{\langle 0| S {\mathcal O}S|0\rangle}{\langle 0|S^2|0\rangle},
\end{gather}
where $S^+=S$ was used and the denominator accounts for the explicit normalization of the many-body wavefunction.

\paragraph{Characteristic function of particle density.}
We focus our attention to the characteristic function of particle density\cite{song,arzamasovs} in a finite spatial interval (or vertex operator\cite{delft}),
given by
\begin{gather}
G_\lambda(x,y)=\left\langle \Psi |\exp\left( 2i\lambda(\phi(x)-\phi(y))\right)\right |\Psi \rangle,
\end{gather}
where  $\frac 1\pi (\phi(x)-\phi(y))=\sum_{n>y}^xc^+_n c_n$ with $x>y$ is the particle number operator within the finite interval from $y$ to $x$ and the equality is
valid within the realm of the low energy theory\cite{song}.
Due to OBC, the system is not translationally invariant therefore the characteristic function depends not only on $x-y$ but independently on the two coordinates.
 
From this, the long wavelength and $2k_F$ oscillating part of the density are obtained as 
\begin{subequations}
\begin{gather}
n_0(x)=\frac{1}{\pi}\lim_{\lambda\rightarrow 0}\partial_x G_\lambda(x,0)/(2i\lambda),\\
n_{2k_f}(x)=G_1(x,0),
\end{gather}
\label{nn}
\end{subequations}
and at half filling and $h=0$, $k_F=\pi/2a$.
By repeating this procedure, any higher moments of the density can be easily evaluated and by Fourier transforming with respect to $\lambda$, 
the distribution function of the density over a finite interval $x-y$ can be obtained for the interacting Hatano-Nelson model.

The characteristic function is evaluated as
\begin{gather}
\ln G_\lambda(x,y)=-2\lambda^2 C(x,y)+\frac{4i\lambda h L}{\pi}\left(g(x)-g(y)\right)
\end{gather}
with 
\begin{subequations}
\begin{gather}
C(x,y)=\left\langle 0\left|\left(\phi(x)-\phi(y)\right)^2\right|0\right\rangle,\\
g(x)=\int_0^L \frac{dx'}{L}  \left\langle 0|\phi(x)\phi(x')|0\right\rangle.
\end{gather}
\end{subequations}
Here, the first term represent the correlation function of the $\phi$ field with open boundary conditions\cite{cazalillaboson}, 
while the second term carries all effect of the non-hermitian term.
These are evaluated to yield
\begin{gather}
\frac 1K C(x,y)=\ln\left(\frac{L}{\pi\alpha}\right)+\frac 12 \ln\left(\sin\left(\frac{\pi x}{L}\right)\sin\left(\frac{\pi y}{L}\right)\right)+\nonumber\\
+\ln\left(\sin\left(\frac{\pi|x-y|}{2L}\right)\right)-\ln\left(\sin\left(\frac{\pi(x+y)}{2L}\right)\right),\label{cxy}\\
g(x)=\frac{K}{\pi} ~\textmd{Im}\sum_{\beta=\pm}\beta ~\textmd{polylog}\left(2, \beta \exp\left(\frac{i\pi x}{L}\right)\right),\label{gx}
\end{gather}
where $\alpha$ is the short distance cutoff in the theory, a remnant of the lattice constant in the $a\rightarrow 0$ continuum limit and
polylog$(2,x)$ is the 2nd order polylogarithm\cite{gradstein}. This is valid in the scaling limit when $L\gg [(x,y,x+y,|x-y|$ mod $L]\gg\alpha$.
This immediately
allows us to
obtain the long wavelength part of the density profile as
\begin{gather}
n_0(x)=-\frac{2Kh}{\pi^2}\ln\left(\tan\left(\frac{\pi x}{2L}\right)\right)
\label{n0}
\end{gather}
on top of the homogeneous particle background. We note that obtaining this result already in the non-interacting, $K=1$ limit is far from being trivial due to the 
non-orthogonality of the single particle eigenfunctions.
For repulsive interactions ($K<1$), the profile flattens out as the particles repel each other, while for the attractive case ($K>1$), the inhomogeneous profile becomes
more prominent as a remnant of the non-hermitian skin-effect.
Interestingly, the very same low energy effective theory applies not only to fermions but to repulsively 
interacting bosons as well, which then do not condense to one end of the sample but
produce a smooth density profile due to repulsion.

\paragraph{Friedel oscillations.}
On top of the long wavelength part, there is also a contribution oscillating\cite{Friedel1952,friedelgravity} with $2k_F$. For this, we need
$C(x,0)=\frac K2 \ln\left(\frac{L}{\pi\alpha}\sin\left(\frac{\pi x}{L}\right)\right)$
and $g(0)=0$. Putting everything together using Eqs. \eqref{nn}, we find the total particle density as
\begin{gather}
\rho(x)=\rho_0+n_0(x)+\nonumber\\
+c\left(\frac{\pi\alpha}{L\sin\left(\frac{\pi x}{L}\right)}\right)^K\sin\left(2k_Fx+\frac{4hL}{\pi}g(x)+\delta\right),
\label{totaln}
\end{gather}
where $\rho_0$ represents the homogeneous background, and 
the coefficients $c$ and $\delta$ are model dependent\cite{cazalillaboson} and cannot be determined from the low energy theory.
For $h=0$, there are no Friedel oscillations at half filling since $k_F=\pi/2a$ makes the oscillations vanish for $\delta=0.$
Eq. \eqref{totaln} represents one of our most important results which indicate that a) the homogeneous density profile gets modified by the imaginary vector potential, b) the spatial
decay of the oscillating term remains intact compared to the hermitian case and c) the oscillation frequency picks up an anomalous term through $hg(x)$.

Close to the middle of the chain, $g(1/2)=2GK/\pi$ with $G\approx 0.916$ the Catalan's constant, and  the oscillation frequency is also modulated
by the imaginary vector potential. Close to the end of the chain, we obtain an effective $x$ dependent wavenumber, summarized as
\begin{gather}
k_F(x)=k_F+\frac{2}{\pi}hK\times\left\{\begin{array}{ll}
\frac{4}{\pi}G, & x\simeq L/2,\\
\ln\left(\frac{2eL}{x\pi}\right), & x\ll L/2.
\end{array}\right.
\label{kfx}
\end{gather}

\begin{figure}[t!]
\centering
\includegraphics[width=8.5cm]{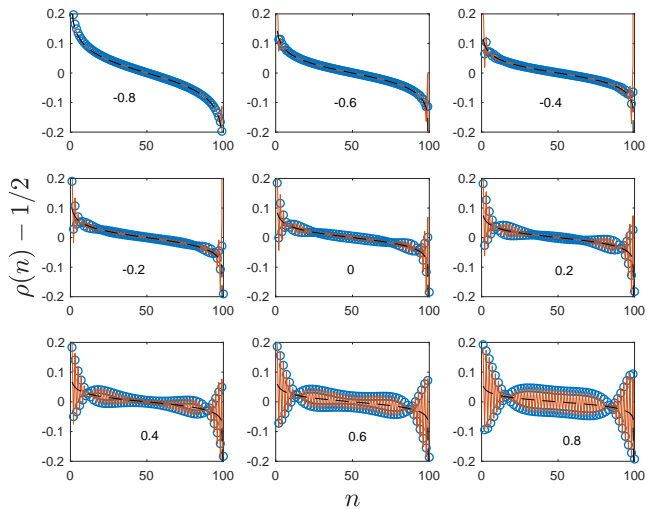}
\caption{Real space density profile and Friedel oscillations for $U/J$=-0.8:0.2:0.8 as indicated in the panels, 
$N=100$ and OBC with $ha=0.1$. The open circles denote
the numerical data from DMRG, while the red solid line represent bosonization from Eq. \eqref{totaln} with $c=1/2$, $\delta=0$ and $L=N\alpha$. The black
dashed line depicts the long wavelength contribution from Eq. \eqref{n0}.}
\label{friedel}
\end{figure}

We compare the prediction of Eqs. \eqref{totaln} and \eqref{kfx} to the numerics using exact diagonalization (ED) and density-matrix renormalization group (DMRG)~\cite{White-1992} calculations of Eq. 
\eqref{hamiltontb} at half filling. For the hermitian case with $h=0$, the particle density is homogeneous and is fixed to $\rho_0=1/2$ even with OBC and interactions. 
In the presence of imaginary vector potential, we still use the exact value of the LL parameter $K=\pi/2/(\pi-\arccos(U/J))$ for small $h$. In Fig. \ref{friedel},
we use $c=1/2$ and $\delta=0$ for all data as the only free parameters and obtain very good agreement with numerics with $L=N\alpha$.
The effect of the modulated Fermi wavenumber from Eq. \eqref{kfx} is visible in Fig. \ref{friedel}, especially close to the boundary.

\paragraph{Probability distribution of the density.}
By the $\lambda$ dependence of the characteristic function, it is apparent that it corresponds to the normal distribution with mean ($\mu$) and variance ($\sigma^2$)
given by
\begin{gather}
\mu=\frac{2 h L}{\pi^2}\left(g(x)-g(y)\right), \hspace*{4mm}\sigma^2= \frac{1}{\pi^2}C(x,y).
\end{gather}
From Eqs. \eqref{cxy} and \eqref{gx}, both the mean and variance scales with $K$ and are enhanced/suppressed for attractive/repulsive interactions.
Moreover, the mean scales linearly with the imaginary vector potential while the variance is insensitive to it within the validity of bosonization.
We emphasize that the variance remains symmetric under the $(x,y)\longleftrightarrow (L -x,L -y)$ transformation, namely the fluctuations are
 insensitive to which end of the system
we consider, while the skin effect in Fig. \ref{hnskin} clearly distinguishes the two ends of the chain within the realm of single particle
physics.

\begin{table}[h!]
\begin{tabular}{c|c|c}
\hline
region & $\mu/hK$ & $\sigma^2 \pi^2/K$\\
\hline
\hline
$x,y\simeq \frac L2$ & $\frac{1}{\pi L}(x-y)(L-x-y)$ & $\ln\left(\frac{|x-y|}{2\alpha}\right)$ \\
\hline
$x,y\ll L$ & $\frac{2}{\pi^2}\left(x\ln\left(\frac{2eL}{x\pi}\right)-y\ln\left(\frac{2eL}{y\pi}\right)\right)$ & $\ln\left(\frac{\sqrt{xy}|x-y|}{\alpha (x+y)}\right)$\\
\hline
$(x,y)=(\frac L2,0)$ & $\frac{4G  L}{\pi^3}$ & $\frac{1}{2} \ln\left(\frac{L}{\pi\alpha}\right)$\\
\hline
\end{tabular}
\caption{Parameters of the normal distribution for various spatial intervals.}
\label{table1}
\end{table}
 
In the followings, we analyze their behaviour for some relevant spatial range.
Close to the middle of the chain with $x,y\simeq L/2$, boundary effects are the most negligible as depicted in Table \ref{table1},
where the variance depends only on $x-y$ and agrees with Ref. \onlinecite{song}, while the mean value is negligibly small.
On the other hand, close to the boundary of the system, the dependence on $x$ and $y$ independently from each other becomes more prominent as summarized in Table \ref{table1}.
Finally, we also consider the case with $y=0$ and $x=L/2$ in Table \ref{table1}, i.e. the distribution of particles in the first half of the chain.

\begin{figure}[t!]
\centering
\includegraphics[width=8cm]{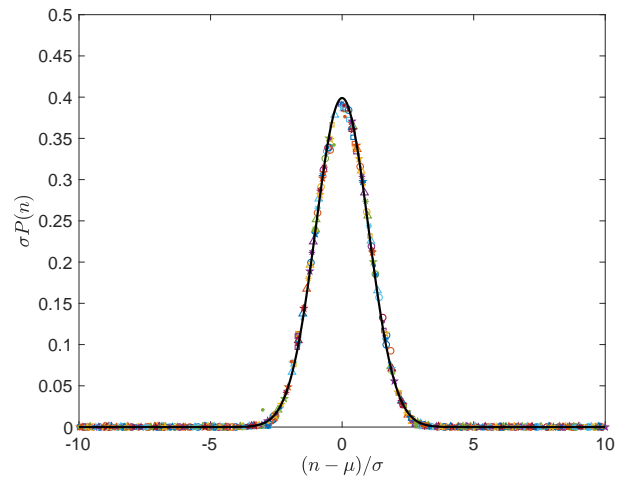}
\caption{Probability distribution of the density of particles in the first half of the chain ($N/2$ sites) from ED and DMRG for $U/J=-0.8:0.2:0.8$ and 
for $ah$=0.1 and $N=28$ (circle), 26 (square), for 
$ah=0.2$ and $N=28$ (triangle), $24$ (star) and 80 (pentagram) 
and for $ah=0.3$ and $N=80$ (dot). The black solid line denotes the standard normal distribution.}
\label{densitydist}
\end{figure}

In order to test the Gaussian nature of the particle distribution, we analyze  the distribution of particles in the first half of the chain, namely
$N_1=\sum_{n=1}^{N/2}c^+_n c_n$, which can take integer values from 0 to $N/2$.
Using ED and DMRG, we evaluated numerically the characteristic function of the particle density, $\langle \Psi|\exp(i\lambda N_1)|\Psi\rangle$
 and after Fourier transforming with respect to $\lambda$, the probability distribution of particle density is obtained. We plot this for several interaction strength, ranging
from strongly attractive through non-interacting to strongly repulsive and for several $h$ in Fig. \ref{densitydist}, after rescaling it with the mean and the variance.
All data fall onto a universal curve, dictated by the standard normal distribution $P(n)=\exp(-n^2/2)/\sqrt{2\pi}$. Already for relatively small system sizes,
the data collapse is excellent, confirming the prediction and validity of bosonization not only for simple expectation value but also for the
full distribution function in non-hermitian systems.

\paragraph{Conclusions.}
We have studied the full counting statistics of particle density in the interacting Hatano-Nelson model. While the non-hermitian skin-effects localizes the single particle
states to one end of the chain, the many-body density profile becomes smooth along the chain, indicating the suppression of the skin-effect by many-body physics.
Friedel oscillations appear throughout the system with beating pattern, which arises from spatially dependent Fermi wavevector due to the imaginary vector potential.
 The probability distribution of particles over any finite interval is found to be normal with the mean scaling
with the imaginary vector potential in spite of the non-hermitian skin effect, which suppresses exponentially the single particle eigenfunctions in one end of the chain.

Our results apply not only to interacting fermions in the Hatano-Nelson model, but 
the very same low energy effective theory through Eqs. \eqref{hboson} and \eqref{thetax} accounts for repulsively
interacting bosons\cite{giamarchi} as well.
Our findings indicate that some peculiar features of single particle non-hermitian physics can be washed out in the many-body realm
and non-hermitian many-body systems behave rather similarly to their hermitian counterparts not only at the level of simple expectation values but also
for the full counting statistics.

\begin{acknowledgments}
This research is supported by the National Research, Development and 
Innovation Office - NKFIH  within the Quantum Technology National Excellence 
Program (Project No.~2017-1.2.1-NKP-2017-00001), K134437, by the BME-Nanotechnology 
FIKP grant (BME FIKP-NAT), and by a grant of the Ministry of Research, Innovation and
 Digitization, CNCS/CCCDI-UEFISCDI, under projects number PN-III-P4-ID-PCE-2020-0277.
\end{acknowledgments}

\bibliographystyle{apsrev}
\bibliography{wboson1}

\end{document}